\documentclass[12pt]{article}

 \usepackage{amsfonts}
 \usepackage{amssymb}
 \usepackage{graphicx} \usepackage{epstopdf}
 \newcommand{\crlb}[1]{\label{#1}\\[2pt]} 
  
 \newcommand{\crld}[1]{\label{#1}}
 \newcommand{\eela}[1]{\quad\hbox{\scriptsize{#1}}\label{#1}\end{eqnarray}}
 \newcommand{\eelb}[1]{\label{#1}\end{eqnarray}}
 
 \newcommand{\newsecb}[2]{\section{#1}\label{#2}\setcounter{equation}{0}}

 \newcommand{\nolabels} {\def\eel{\eelb} \def\crl{\crlb} \def\newsecl{\newsecb}\def\bibiteml{\bibitem}\def\citel{\cite}\def\labell{\crld}}

\newcommand\publishversion{\nolabels\setlength{\textheight}{8.6in}\setlength{\oddsidemargin}{0in}
    \setlength{\textwidth}{6.3in}\setlength{\topmargin}{-0.3in}}

                 \def\fn{\footnote}
        \def\be{\begin{eqnarray}}    \def\ee{\end{eqnarray}}
 \def\bi#1{\begin{itemize}\item[#1]}   \def\itm#1{\item[#1]}  \def\ei{\end{itemize}}  
   \def\^#1{\hat{#1}}

         \def\F{\Phi}

      \def\ket{\rangle}

 \def\iss{\ =\ }

 \def\ffract#1#2{\raise .2 em\hbox{$\scriptstyle#1$}\kern-.3em/
                 \kern-.2em\lower .15 em \hbox{$\scriptstyle#2$}}

\def\bmatrix{\begin{matrix}} \def\ematrix{\end{matrix}} \def\bpmatrix{\begin{pmatrix}}\def\epmatrix{\end{pmatrix}}
\def\bcenter{\begin{center}} \def\ecenter{\end{center}}

\def\lowerheightfig#1#2#3{\(\raise-#1\hbox{\includegraphics[height=#2]{#3}}\)}
\def\lowerwidthfig#1#2#3{\(\raise-#1\hbox{\includegraphics[width=#2]{#3}}\)}

\publishversion

\begin{document}
\bcenter 
{ \LARGE\textbf{ Free Will in the Theory of Everything\fn
{Presented at the Workshop on ``Determinism and Free Will", Milano,  May 13, 2017.} \\[25pt] }}
{\large Gerard 't~Hooft}  \\[20pt]
Institute for Theoretical Physics \\[5pt]
\(\mathrm{EMME}\F\)   \\
Centre for Extreme Matter and Emergent Phenomena\\[5pt] 
Science Faculty\\ 
Utrecht University\\[5pt]
 POBox 80.089 \\
 3808 TB, Utrecht  \\
The Netherlands  \\[10pt] 
e-mail:  g.thooft@uu.nl 
\\ internet:  http://www.staff.science.uu.nl/\~{}hooft101/ 
\ecenter
\vfil
 \noindent {\large\bf Abstract }
 
 	From what is known today about the elementary particles of matter, and the forces that control their behaviour, it may be observed that still a host of obstacles must be overcome that are standing in the way of further progress of our understanding. Most researchers conclude that drastically new concepts must be investigated, new starting points are needed, older structures and theories, in spite of their successes, will have to be overthrown, and new, superintelligent questions will have to be asked and investigated. In short, they say that we shall need new physics.
	
	Here, we argue in a different manner. Today, no prototype, or toy model, of any so-called Theory of Everything exists, because the demands required of such a theory appear to be conflicting. The demands that we propose include locality, special and general relativity, together with a fundamental finiteness not only of the forces and amplitudes, but also of the set of Nature's dynamical variables. We claim that the two remaining ingredients that we have today, Quantum Field Theory and General Relativity, indeed are coming a long way towards satisfying such elementary requirements. Putting everything together in a Grand Synthesis is like solving a gigantic puzzle. We argue that we need the correct analytical tools to solve this puzzle. Finally, it seems to be obvious that this solution will give room neither for ``Divine Intervention", nor for ``Free Will", an observation that, all by itself, can be used as a clue. We claim that this reflects on our understanding of the deeper logic underlying quantum mechanics. \vfil 
\noindent Version September  2017	\hfill		Typeset \today

\eject
\setcounter{page}{2} 

\newsecl{Theories of Everything}{everything}
What is a ``Theory of Everything"? When physicists use this term, we begin with emphasising that this should not be taken in a literary sense. It would be highly preposterous for any domain of science to claim that it can lead to formalisms that explain ``everything". When we use this phrase, we have a deductive chain of exposition in mind, implying that there are `fundamental' laws describing space, time, matter, forces and dynamics at the tiniest conceivable distance scale. Using advanced mathematics, these laws prescribe how elementary particles behave, how they exchange energy, momentum and charges, and how they bind together to form larger structures, such as atoms, molecules, solids, liquids and gases. The laws have the potential to explain the basic features of nuclear physics, of astrophysics, cosmology and material sciences. With statistical methods they explain the basis of thermodynamics and more. Further logical chains of reasoning connect this knowledge to chemistry, the life sciences and so on. Somewhat respectlessly, some might try to suggest that a `theory of everything' lies at the basis of most of the other sciences, while such of course is not the case.  The impression must be avoided that other sciences would be thought to be less `fundamental'. In practice, a theory of everything would not affect much of the rest of science at all, simply because each of the elements of such a deductive chain would be far too complex and far too poorly understood to be of any practical value. The theory applies to `everything' only in a formal sense.

What do physicists imagine a ``Theory of Everything" to look like? Should it be a `Grand Unified Theory' of all particles and forces? If so, we are still a far way off, since the relevant distance scale at which fundamental modifications of our present theoretical views are expected to be needed is the so-called Planck Length, some \(10^{-33}\) cm, which is more than a billion times a billion times smaller than anything that can be studied directly with laboratory experiments. Is it `quantised gravity'? There are deep and fundamental problems that arise when we try to apply the principles of quantum mechanics to the gravitational force. Forces and quantum mechanical amplitudes tend to infinity, and the remedies for that, as proposed up to today, still seem to be very primitive. Since they lead us out of the perturbative regime, calculations are imprecise, and accurate definitions explaining what we are talking about are still missing. 

Is it `Superstring Theory'? The problem here is that this theory hinges largely on `conjectures'. Typically, it is not understood how most of these conjectures should be proven, and many researchers are more interested in producing more, new conjectures rather than proving old ones, as this seems to be all but impossible. When trying to do so, one discovers that the logical basis of such theories is still quite weak. One often hears the argument that although we do not quite understand the theory `yet', the theory is so smart, that it does understand how it works itself. And then: its mathematics is so beautiful and coherent, this theory `must be true'. To this author's mind, such arguments are more dubious than often realised, if the history of science can be used as a clue.

Finally, many researchers are tempted to part from the established paths to try `completely new and different' starting points. Often, these are not based on sound reasoning and healthy philosophies, and the chances for success appear to be minimal.
Perhaps the reader's first impression of this paper would not show it, but this paper is a plea for rigorous reasoning and accurately keeping established scientific results in mind.

	Is humanity smart enough to fathom the complexities of the laws of Nature? If history can serve as a clue, the answer is: perhaps; we are equipped with brains that have evolved a little bit since we descended from the apes,  hardly more than a million years ago, and we have managed to unravel some of Nature's secrets way beyond what is needed to build our houses, hunt for food, fight off our enemies and copulate. 
In terms of cosmic time units, a million years is not much, and up to today, our brains have not evolved for carrying out this particular new task, but we may just about manage to figure things out, making numerous  mistakes on our way. The nice thing about science is that mistakes can be corrected, so we are standing a chance.
	
	Today's attempts at formulating ``theories of everything" must look extremely clumsy in the eyes of beings whose brains have had more time, say another few million years, to evolve further. This author is convinced that many of the starting points researchers have investigated up to today, are totally inappropriate, but that cannot be helped. We are just baboons who have only barely arrived at the scene of science. Using my own limited brain power, I am proposing a somewhat different starting point.
	
	The following two sections, the main body of the lecture given, may look peculiar, contemplating the laws of nature from an unconventional vantage point. We argue that the fundamental laws of nature appear to be chosen in an extremely efficient way. The only thing that may seem \emph{not} to agree with our philosophy, is quantum mechanics. On the other hand, also quantum mechanics does appear to be an extremely efficient theory. Without quantum mechanics, we would not have been able to construct meaningful theories for atoms and sub-atomic particles at all.
	
	The good thing about quantum mechanics is the simple fact that many of nature's variables that used to take continuously varying values in classical physics, now turn out to be quantised. In Section \ref{quantum} we summarise observations concerning the mathematical coherence of quantum mechanics. One could ascribe the very special logical structure of quantum mechanics to the inherent discreteness of the physical variables it describes. Now this special form of logic also seems to force us to abandon the notion of definiteness of observables, as if nothing can be absolutely certain in a quantum system. But looking deeper in the mathematical structure of the theory, one can question such conclusions.  The author has somewhat deviating views on quantum mechanics, which we briefly explain in Section \ref{bell}. 
	
	Our conclusion will be that our world may well be super-deterministic, so that, in a formal sense, free will and divine intervention are both outlawed. We emphasise however that, in daily life, nobody will suffer from the consequences of such an observation; it pertains to the deeper fundamental nature of physical laws.

\newsecl{God's assignment}{God}
	Imagine that you were God\fn{This is only meant metaphorically; this author, fortunately, is not religious.} . Your assignment is: \emph{run a universe}. Your universe may look like a big aquarium, containing things like stars and planets, plants, animals, humans, and more elementary objects such as atoms and sub-atomic particles. To make it all work, you will want billions or more of all of these.
You may steer all these objects in any way you like, and you want interesting things to happen. What would you do?

	You would have a problem. To tell every individual object in your universe what to do will require a massive amount of administration. Suppose you want to be efficient, isn't there an easier way? The answer is yes. You declare that there are rules. Every object, every particle this object is made of, moves around as directed by \emph{laws of Nature}. Now, there are only two things left to be done: design laws of Nature, and attain a powerful computer to help you implement the laws of Nature. Let us assume that you have such a powerful computer.  Then the question  is: how do you choose the laws of Nature? 
	
	Stars, planets and people are quite complex, so you do not want the rules to be too simple, since then nothing of interest will happen in your universe. Computer scientists would have ideas about designing rules, a software routine, a program, telling you how your universe evolves, depending on the laws you feed it with, but to make your universe sufficiently realistic, their programs will tend to become lengthy, complex, and ugly. You want to be more demanding. 
	
	So, being God,  you have a second great idea. Before formulating your laws of Nature, you decide about a couple of \emph{demands} that you impose upon your laws of Nature. Tell your computer scientists and mathematicians that they must give you the simplest laws of Nature that comply with your demands.
	
	While listening to what your computer scientists and mathematicians tell you about the viability of your demands, you copy the rules that they formulate.  You impose the rules, and you press the button.

	In this paper, it will be argued that very simple demands can be imposed, and that at least some of these demands already lead to a structure that may well resemble our universe. The construction that will eventually emerge will be called "theory of everything". It describes everything that happens in this universe. 
Now, the first demand that will be suggested here, will appear to be not al all obeyed by the real universe, at first sight. But those are only appearances. Remember that our brains were not designed for this, so wait with your prejudices. We claim to be able to make three observations: one, the set of demands  that we will formulate now are nearly inevitable and non negotiable. Two: even though the demands are simple, the mathematical structure of the rules, or laws of physics, emerges to be remarkably complex, possibly too complex for simple humans to grasp. And three: as far as we do understand them,
the resulting rules do resemble the laws of Nature governing our real universe. In fact, it may well be that they lead \emph{exactly} to our universe. This is my projected path towards a ``Theory of Everything".

\newsecl{Demands and rules}{dems}

\begin{quote}\textbf{Demand \# 1:}  Our rules must be unambiguous. At every instant, the rules lead to one single, unambiguous prescription as to what will happen next. \end{quote}

\noindent Here, most physicists will already object: \emph{What about quantum mechanics?} Our favoured theory for the sub-atomic, atomic and molecular interactions dictates that these respond according to chance. The probabilities are dictated precisely by the theory, but there is no single, unambiguous response.

I have three points to be made here. One: This would be a natural demand for our God. As soon as He admits ambiguities in the prescribed motion, he would be thrown back to the position where gigantic amounts of administration is needed: what will be the `actual' events when particles collide? Or alternatively, this God would have to do the administration for infinitely many universes all at once. This would be extremely inefficient, and when you think of it, quite unnecessary. This God would strongly prefer one single outcome for any of His calculations. This, by the way, would also entail that his computer will have to be a \emph{classical} computer, not a quantum computer, see Refs.\,\cite{Zuse,fredkin,feinstein}.

Second point: look at the universe we live in. The ambiguities we have are in the theoretical predictions as to what happens when particles collide. What \emph{actually} happens is that every particle involved chooses exactly one path. So God's administrator must be using a rule for making up His mind when subatomic particles collide.

Third point: There are ways around this problem. Mathematically, it is quite conceivable that a theory exists that underlies quantum mechanics.\cite{GtHCA} This theory will only allow single, unambiguous outcomes. The only problem is that, at present, we do not know how to calculate these outcomes. I am aware of the large numbers of baboons around me whose brains have arrived at different conclusions: they \emph{proved} that hidden variables do not exist. But the theorems applied in these proofs contain small print. It is not taken into account that the particles and all other objects in our aquarium will tend to be strongly \emph{correlated}. They howl at me that this is `super-determinism', and would lead to `conspiracy'. Yet I see no objections against super-determinism, while `conspiracy' is an ill-defined concept, which only exists in the eyes of the beholder. A few more observations on this topic are made in Section \ref{bell}.

\begin{quote} \textbf{Demand \# 2:} We must have \emph{causality}: every event must have a cause, and these causes all must lie in the past, not in the future. \end{quote}

\noindent A demand of this sort is mandatory. What it really means is that, when our God looks up his rules to figure out what is supposed to happen next, He should never be confronted with a circular situation, or in other words, He should always know in which \emph{order} the rules must be applied. Whatever that order is, can be used to define \emph{time}. So now we can distinguish future from past. Only the past events are relevant for what happens next, and whatever they dictate, will only affect the future. This principle has been instrumental in helping us understand quantum field theories, for instance.

\begin{quote}\textbf{Demand \# 3:} \emph{Efficiency} Not all events in the past, but only a few of them should dictate an event in the presence. \end{quote}

\noindent This suggests that there is a power limitation in God's laptop. We can't have that every particle in the past, instantly determines the behaviour of every particle in the future. If, for computing the behaviour of one particle, we only need the data concerning a few particles in its immediate environment, then the calculation will go a lot quicker. We are simply asking for a maximum of efficiency; there are still a lot of calculations to do.

This demand will now also lead to our first rule, or law of physics:

\begin{quote}\textbf{Rule \#1:} \emph{Locality}.  All configurations one needs to know to determine the behaviour of an object at a given spot, must be lying in its vicinity. \end{quote}

\noindent This means that we will have to define \emph{distances}. Only points at very small distances from a given point are relevant for what happens there. What `vicinity' really means still has to be defined. In our universe it is defined by stating that space is three-dimensional, with a Euclidean definition of distance. Details will be left for later.

Now we still have to decide how things interact, but before that, we have to decide how things can move. This is a delicate subject. In our universe, things that can stand still, will be allowed to move along straight lines, with, at first sight, any velocity. This makes things difficult for God's computer programmer. A programmer will find it easy to define objects that move with a pre-designed velocity in a pre-designed direction, but what are the rules if something may move with \emph{any} speed in just \emph{any} direction? A deceptively simple-looking answer is in the form of a new law of Nature,

\begin{quote} \textbf{Rule \#2:} \emph{Velocity}.  Any object for which it has been decided how it behaves when at rest, will behave very similarly when it moves along a straight line in any direction, with any constant velocity (within limits, see the next rule). \end{quote}

\noindent The rule of its behaviour at this velocity must be derivable in a simple way from what the rules are when it stands still. Think of someone sitting in a train, playing chess. How this person feels, how he moves his arm while moving a pawn, as well as the rules for chess, all are the same when the train moves, as what they are when the train stands still.

This is an important rule, since it adds an enormous amount of structure and complexity in our universe. The relative positions of things can now change with time, but things can also collide, they can have moving parts, etc. At first sight, the price we pay for this added complexity may seem to be mild, since we get moving things for free, if we know how they function when standing still.

But there is a problem. Should we accept \emph{all} velocities, or should there be a speed limit? If you don't impose a speed limit, you run into trouble. The trouble is with locality. If things move infinitely fast, they can be simultaneously here and far away. In practice, this also means that there will be trouble with our demand of locality and/or efficiency. Many of our particles will move so fast that our computer needs infinite processing speed. This we should not allow.   Therefore: 
\begin{quote} \textbf{Rule \#3:} There is a \emph{speed limit}. Call it \(c\), the speed of light. \end{quote}

\noindent This fits well with locality: the only neighbours that interact with a particle of matter, are the ones that can be reached by a light signal within a limited time step.

We know that there is a speed limit in our universe: the speed of light.
Thus, we saved efficiency and locality, but now there is a new problem. The person in the train moves his arm. The train may be moving slower than the speed limit, but what about the arm? Well, His mathematicians tell God that this problem can be solved, but a number of amendments must be made for rules \#2 and \#3. First, let us slow down the arm:

\begin{quote} \textbf{Rule \#3a:} Things that go very fast, will have time going slower. \end{quote}

\noindent This slows down the arm, but it is not quite enough. We also need

\begin{quote} \textbf{Rule \#3b:} Things that go very fast, will also contract in the forward direction. \end{quote}

\noindent This makes the arm shorter, but again it does not help quite enough. A more drastic measure:

\begin{quote} \textbf{Rule \#3c:} Inside things that move fast, clocks will no longer go synchronously. \end{quote}

\noindent This, in combination with rules \#3a and \#3b, works: A person inside the train may stick out his arm, but as seen from outside the train, the arm reaches the pawn at the moment that the body has nearly overtaken the arm. Mathematicians now tell us that we need all three of these amendments to rule \#3, and now the logic works out fine; the arm will not exceed the speed limit.

Physicists have learned about this rule, with its amendments, along somewhat different lines of reasoning, but the result is the same: Einstein's special relativity. As we see here, Einstein's relativity theory could have been deduced from purely logical arguments, if our brains had been hundreds of times smarter. In our world, it was arrived at by Hendrik Antoon Lorentz, by studying the laws of electro-magnetism. We see now why we could never have understood our universe if there hadn't been Special Relativity.

We still haven't tried to determine \emph{how} things behave, even if everything stands still. This is because there is another problem. Think of an object, such as a lump of sugar. It is extended in space. Because of locality, one side of our lump of sugar should behave independently from the other side. Should it not be possible to break the lump of sugar in half? And the pieces we then get, should we not be able to break these in halves again? And so on? Can we break these pieces in halves forever?

This question was already raised by the Greek philosophers Democritus, Leucippus and Epicurus around 400 BC. In a stroke of genius, they stumbled upon the right answer: \emph{No, this series of divisions will stop}. There will be a smallest quantity of sugar. They called these smallest quantities `atoms'. So it was these Greeks who first tried the concept of \emph{quantisation}. They quantised matter, purely by using their brains. Our God is forced to assume something like this as well, since, if things could be divided into pieces ad infinitum, this would imply infinite complexity, which needs to be avoided.

In the name of our efficiency demand, we must quantise as well. All objects in this universe can be broken into smallest units. The `atoms of sugar' are now known as `molecules', but that's just a detail. Molecules were found to be composed of smaller things, and these are now called `atoms', while they can be divided further: the smallest possible objects are now called elementary particles. Note that, in modern theories for elementary particles, these particles are denoted as single points in space. A point cannot be divided in two. When, nevertheless, a particle such as an electron, emits another particle, for instance a photon, then in particle physics we say that the photon is created on the spot; it does not hide inside an electron. In conclusion, it turns out that we need:

\begin{quote} \textbf{Rule \#4:} Matter is \emph{quantised}. The smallest quanta of matter are the elementary particles. \end{quote}

\noindent But for God's  mathematicians, these quanta of matter are causing considerable trouble. The quanta will probably carry mass, energy and momentum, but even if they are point-like, we sometimes do need a property replacing the notion of size. It was found how to do this: for all particles with finite amounts of momentum, there is a natural smallest size limit. The math needed is called quantum mechanics.

Quantum mechanics works, but it is complicated. Yet, up to this point, today's physicists and mathematicians have discovered how to combine these rules in a working configuration. In particular, adding special relativity, rule \#3, took us nearly 50 years, so it wasn't easy. The result was called ``quantum field theory". There is one problem with quantum field theory: it is not known where the \emph{forces} come from; this leaves us with lots of freedom, as it is not known how God made His decisions here.

Thus, we have to introduce one more concept: forces. In our universe, it must be possible to \emph{change} the velocities of objects, and decide about a rule for this, of the following type:

\begin{quote}	\textbf{Rule \#5:} \emph{Forces.} If it is known how an object behaves while moving on a straight line with constant velocity, it should be possible to deduce how it behaves moving with a varying velocity on a curved line. \end{quote}

The primary force that can be deduced this way is gravity. In our world, we know that other forces exist, but these may be due to secondary effects resulting from complex behaviour at ultrashort distances.
Again, there will be a price to pay: the best way to add the notion of curved lines in the logic of our rules, is to have curvature in the fabric of space-time itself. One then may create the situation that the fundamental differences between straight lines and curved lines disappear; on curved spaces, straight lines do not exist.

There is also an other advantage: curved universes have no fixed size, they can expand. This means that our universe may begin by being very tiny and very simple, and grow all by itself, just by the action of our rules.
In our universe, this situation occurs. The theory describing these aspects is Einstein's theory of General Relativity. 

This leads us to one more rule:

\begin{quote} \textbf{Rule \#6:} God must tell his computer what the \emph{initial state} is. \end{quote}

\noindent Again, efficiency and simplicity will demand that the simplest possible choice is made here. This is an example of Occam's rule. Perhaps the simplest possible initial state is a single particle inside an infinitesimally small universe. 

Final step:

\begin{quote} \textbf{Rule \#7:} Combine all these rules into one computer program to calculate how this universe evolves. \end{quote}

\noindent So we're done. God's work is finished. Just push the button. However, we reached a level where our monkey brains are at a loss. Rules \#5 -- \#7 have shown to be too difficult for us. The theory of general relativity manages  to take Rule \#5 into account, but unfortunately does not handle quantum mechanics, rule \#4, correctly.

Why is this so difficult? Quite possibly, more rules have to be invented to reach a coherent evolution law, but as yet, the question is staring us in the face. Can we implement all rules given above into a single, working scheme? What comes out will be of secondary importance. Perhaps a framework will be found with many possibilities (a ``multiverse"). In that case, more rules have to be invented to single out one preferred choice. So-far, it seems that the requirements we did mention above have all been taken in consideration in our universe's laws of physics. This is why this author suspects that the given rules make a lot of sense.

We note that the actual laws of physics known to hold in our universe are quite close to what we constructed purely by mental considerations. Of course, the author admits that this will be attributed to hind-sight, but we claim that a super intelligent entity could perhaps have ``guessed" nature's laws of physics from such first principles. This would be important to know, since this is an encouragement to use similar guesses to figure out how the remainder of physical laws, not yet known to us today, might be guessed as well.

 The idea of underlying laws that are completely mechanical, while quantum mechanics, as it is known, should be an emergent feature of the universe, has been suggested several times\,\cite{Zuse,fredkin, feinstein}, but there are deep problems with it, which will be addressed now.

\newsecl{Free Will}{freewill}

Note what the motivations for the demands formulated in Section \ref{dems} have been: unambiguity, simplicity, efficiency and finiteness. In particular this last demand, finiteness, is not (yet?) completely implemented in the known laws of Nature today. There are various things that can go out of control due to infinities. In quantum field theories, we managed to keep one kind of infinities under control, the infinities in the quantum amplitudes and all physical effects associated to those. This means that the effects of forces in the theory stay finite and computable. 

This is important. However, we always need to restrict ourselves to approximations, \emph{in casu} using perturbation expansion techniques. An infinity that we left aside because, superficially, it did not harm us, is the infinity of all physical dynamical variables that participate. For us this seems to cause no problems, as long as our integrals converge, but for a `God' who wishes to stay under control of everything, this is not an option: the total number of independent variables must be finite. His Laptop must be able to compute \emph{exactly} what happens in a finite stretch of time. Here, our arguments seem to favour a universe that is spatially compact rather than unbounded.

The reader might accuse me of an ill-motivated religious standpoint, but we do note that the rules we arrived at by using this standpoint are remarkably effective in generating laws of Nature that are known to work quite well.

Then, the reader might bring forward that no classical laptop at all can compute quantum mechanical amplitudes with infinite precision, a \emph{quantum}-laptop would be needed.

This however, might be the result of an elementary incompleteness in our present understanding of quantum mechanics, as we argued some time ago\cite{GtHCA}. There is every reason to suspect that a novel theory underlying quantum mechanics will be required. To satisfy our demand of unambiguity, all phenomena must be entirely computable, not left to chance. 

If this is right, the laws of Nature we arrive at leave no room for two things:
\bi {-} Divine Intervention, and
\itm{-} Free Will. \ei
We claim that there would be very little justification for the existence of either. If we would allow for Divine Intervention, for instance in all quantum mechanical phenomena, our theories would leave such a gigantic amount of arbitrariness in its prescriptions, that all laws of physics would seem to be there for nothing. We would find ourselves back on square one.

As for free will, the argument is very similar. If quantum mechanics would leave space for free will, there would be far too much space for that. There is every reason to suspect that today's voids in the theory of quantum mechanics will be filled by additional laws.

Most importantly, quantum mechanics itself can be used to show to us how the voids might be filled in. It is not hard to imagine versions of our dynamical theories where what looks like quantum mechanics today can be aptly described in classical terminology, but we need these missing laws.

It is important to note that quantum mechanics accurately predicts the statistics when experiments are repeated many times. If there are additional laws that decide about individual events, these laws must reproduce the statistics as it is predicted by quantum mechanics alone. This implies that the question whether the additional laws exist or not will not be decidable experimentally. Physicists who are content with a theory that never gives better answers than statistical ones, will categorically reject speculations concerning hidden variables, but religious people who assume that our universe is reigned by some God, should require that quantum mechanics  be complemented with theories of evolving hidden degrees of freedom, in such a way that all events that take place can be attributed to something that has happened nearby in the past.

Whether Devine Intervention takes place or not, and whether our actions are controlled by ``free will" or not, will never be decidable in practice. This author suggests that, where we succeeded in guessing the reasons for many of Nature's laws, we may well assume that the remaining laws, to be discovered in the near or distant future, will also be found to  agree with similar fundamental demands. Thus, the suspicion of the absence of free will can be used to guess how to make the next step in our science.

\newsecl{Quantum Mechanics}{quantum} Today's scientists have not yet reached that point.  All dynamical laws in the world of the (sub)atomic particles were found to be controlled by  quantum mechanics.  Quantum mechanics appears to add a sense of `uncertainty' to all dynamical variables describing these particles: you can't have position and momenta of particles be sharply defined at the same time, components of the spin vector contain a similar notion of uncertainty, and sometimes the creation of a particle can be confused with the annihilation of an antiparticle -- and so on. This is actually not a shortcoming of the theory, because in spite of these apparent uncertainties, statistical properties of the elementary particles can be determined very precisely. What is going on?
	
	The rules according to which quantum mechanics works, are precisely formulated in what is sometimes called the Copenhagen interpretation. This is not the place to explain precisely what the Copenhagen rules are, but they can be summarised by stating that the behaviour of a particle can be described as completely as if there were no `uncertainties' at all. Instead, we have variables that do not commute:
	\be x\cdot p-p\cdot x\iss[\,x,\,p\,]\iss i\,\hbar\ , \eel{xpcomm}
In practice, what this means is that, when a system is described quantum mechanically, we apply a number system that is more general than in classical physics, but just as applicable. In fact, this number system is more useful than the old, commuting numbers, in case it refers to quantities that are quantised, that is, they only come in integer multiples of fixed packages.

	Thus, imagine a system that allows its dynamical variables only to occur in distinct states, typically indicated by integer numbers, \(|1\ket,\ |2\ket,\ |3\ket,\ \dots\). The non-commuting numbers that we use can then be called \emph{observables} in case they simply describe the state the system is in, like indicating the value that a particular integer has. In case they are applied to replace a state by an other state,  they are called \emph{operators}; for instance, \(a|n\ket=|n-1\ket\). The manipulations with which we handle these numbers act the same way regardless whether we are dealing with observables or operators. This makes quantum mechanics extremely flexible, but it sometimes obscures the situation when we are unable to distinguish observables from operators.
	
	A \emph{quantum transformation} replaces observables by operators, or more often, mixes the two types completely. One ends up having to use wave functions to describe the states a system can be in. The beauty of this formalism is that such numbers, the non-commuting numbers, regain continuity even if the original system was discrete, and as such allow us to use the machinery of advanced mathematics. This leads to such powerful results that only few physicists are ready to return to the original system of discrete physical states describing `reality'. After most quantum transformations, reality is replaced by the more abstract notion of a wave function. This notion only appears to be abstract if one asks ``what is going on here?", but in practice serves us very well if we only ask ``What will the result of this experiment be?"
	
	In fact, according to the Copenhagen interpretation, questions such as ``What is going on here?" are ill-posed questions, as they cannot be answered by doing experiments. In practice, therefore, we usually refrain from asking such questions. All that matters is the reproduction of the answers given by experiments.
	
	Nevertheless, our question ``What is gong on here?" is not ill-posed. We can always attempt to find answers of principle: we do not know what is going on here, but we can imagine very precisely what it could be. \emph{Something} is going on, and the assumption that there is something going on that might explain what is happening next, even if we cannot be certain what it was, may be used as an important constraint in constructing theories. A typical example is the Standard Model of the sub-atomic particles. This model was established partly by doing experiments with elementary particles, but also by imagining ``How should these particles behave?" It makes sense to use as an assumption: every particle behaves in a completely deterministic way, even though its behaviour cannot be determined completely by any known observation technique. The assumption that particles behave in such a way that a completely deterministic theory is responsible is not a crazy assumption, but it requires guesswork. In principle, such guesses could help us to guess correctly what the next stage for the Standard Model might be.

\newsecl{Bell's theorem}{bell}
The reasoning summarised above, which we explained more elaborately in Ref.\,\cite{GtHCA}, may seem to be logical, yet it is nearly universally rejected by researchers in quantum mechanics. The reason for that is that there seems to exist a rigorous proof of the contrary statement: \emph{experiments can be carried out, for which standard quantum theory provides very firm predictions concerning their outcomes, predictions that have indeed been confirmed by experiment, while they do not allow for any `ontological' description at all.} The question ``What is going on here?" cannot be answered at all without running into apparent contradictions.

For a complete description of Bell's Gedanken experiment, we refer to the literature, and references therein\,\cite{bell,CHSH-1969}. Here, we summarise.

Bell's starting point is, that experimenters can put small particles in any quantum state they like, and this appears to be true in most cases. In particular, we can put a pair of particles in an \emph{entangled} quantum state. Photons, for example, are described not only by a plane wave that determines in which direction a photon goes, but also by its polarisation state. Photons can be linearly polarised or circularly polarised, but there are always exactly two possibilities for the polarisation: vertically or horizontally, or alternatively, left or right circularly polarised.

An atom can be put in an excited state in such a way that it emits two photons, which, together, form only one possible quantum state: if one photon is found to be vertically polarised, the other will necessarily be vertically polarised as well, and if one photon is circularly polarised to the left, the other is polarised to the left as well. In this case, the two photons form an {entangled} state. Together, however, this is only one single, allowed quantum state that the pair of photons can be in.

Far from the decaying atom, Bell now imagines two detectors, called Alice and Bob, each monitoring one of the photons. They both work with linear polarisation filters, checking the polarisation of the photon that they found. They do  a series of experiments, and afterwards compare their results. They do not disclose in advance how they will rotate their polarisation filters. Now, whenever the two polarisation filters happen to be aligned, it turns out that they both measure the same polarisation of their photons. When the two polarisation filters form an angle of \(45^\circ\), they find the two photons to be totally uncorrelated. But when the relative angle is \(22.5^\circ\) or \(67.5^\circ\), they find a relatively high correlation of the two polarisations. In classical physics, no model can be constructed that reproduces this kind of correlation pattern.

The only way to describe a conceivable model of ``what really happens", is to admit that the two photons emitted by this atom, know in advance what Bob's and Alice's settings will be, or that, when doing the experiment, Bob and/or Alice, know something about the photon or about the other observer. Phrased more precisely, the model asserts that the photon's polarisation is \emph{correlated}\fn{This can also be rephrased as follows: the assumption that, in the initial state, the experimenter can always produce any entangled state of photons as he or she pleases, is not true. There will be strong and uncontrollable\,---\,entangled\,---\,correlations with other atoms in the system.} with the filter settings later to be chosen by Alice and Bob. We can compute what kind of correlation is needed. One finds that the correlation is a pure \emph{three-body correlation}: if we average over all possible polarisations of the photon pair, we find that Alice and Bob are uncorrelated. If we average over all possible settings Alice can choose, then Bob's settings and the polarisation of the photons are again uncorrelated, and \emph{vice versa}. 

But this three-body correlation is said to be impossible. How can the photons know, in avance, what Bob and Alice will do? The required correlation appears to contradict the notion that, after the photons have been emitted, both Alice and Bob have the \emph{free will} to choose any setting they like. Have they?

It is easy to say that they have not. If we adhere to a deterministic model, it is clear that the polarisation of the photon, as well as the settings chosen by Alice and Bob, have been determined by the initial state of the universe, together with deterministic equations of motion. But this is not the complete answer to our problem. How do we make a model for these photons?

Apparently what we have here, is a strong 3-body correlation. The three points in space and time that are correlated, may well all be spatially separated from one another. This means that no signal can have been transmitted from one to the other, but this is not a problem. It is well known in the quantum theory of the sub-atomic particles, that correlations need not vanish outside the light cone.\fn{In contrast, the \emph{commutator} of two operators defined on a pair of space-time pints, does vanish outside the light cone. The commutator can be seen to monitor causal influences of one operator on the value of the other, and so one can derive that a non-vanishing commutator will enable experimenters to send signals to one another, while the \emph{correlation function} only points towards a common past of the pair of space-time points.}  The real problem here is that Alice's and Bob's settings are classical, and the quantised atom was there first. What kind of model can bring about such strong correlations, even if they are 3-point correlations, when two of the variables considered are classical?

If this is the way to look at the problem raised by Bell's theorem, we can limit ourselves to a more elementary question. Consider just a single, polarised photon. It may have been emitted by some quasar, billions of years ago. An observer detects it after it passed a polarisation filter. The photon either passes or it does not. In both cases, the `true polarisation state' of the photon was either in line with the observer's filter, or orthogonal to it, but not in any other direction. It seems as if the quasar, billions of years ago, already knew that these were the two polarisation directions the photon had to choose from. This will be a strange aspect of any model that we might want to apply.

And now for what this author believes to be the correct answer, both for the single photon problem and the Bell experiment.
Our theory is that there does exist a true, ontological state, for all atoms and all photons to be in.  All ontological states form an orthonormal set, the elements of an ontological basis. The universe started out to be in such a state, and its evolution law is such that, \emph{at all times in the future, the universe will still be in an ontological state}. Regardless which ontological initial state we start from, the state in the future will be an ontological one as well, that is, \emph{not a quantum superposition of different ontological states}. What we have here, is a conservation law, the conservation of ontology. It selects out which quantum superpositions can be allowed and which not, just because, according to our model, the evolution law is ontological.

Consider the single photon emitted by our quasar. Regardless of what will happen in the future, the quasar and all photons it emitted, together form an ontological state. The superposition of two ontological states is not ontological, because all ontological states are orthogonal to one another.

The photon that eventually passes the observer's filter, leads to a classical observation, so it is ontological. Any photon that would pass a filter not orthogonal to the previous one, will not be an ontological photon, and, by following our ontology conservation law backwards in time, the quasar can never have emitted such a photon. In short, when our quantum theory is underpinned by an ontological evolution theory, its Schr\"odinger equation will exclusively select ontological states to evolve into. If that rule requires Bob and Alice to choose from a particular subset of possible settings then so be it; this is what the model predicts.

How can our model force the late observer, Alice, or Bob, to choose the correct angles for their polarisation filters? The answer to this question is that we should turn the question around. Suppose any of the late observers choose a different polarisation angle. Only a superposition of the original photons passes this filter with certainty, whereas we know that the early photons are correlated to the late, classical filters. But such a correlation is not strange; no \emph{conspiracy} is required. We must accept that the ontological variables in nature are all strongly correlated, \emph{because they have a common past}. We can only change the filters if we make some modifications in the initial state of the universe. This same modification will necessarily also affect the photons emitted by the atom, so as to comply with the ontology conservation law. Perhaps one could call this ``conspiracy", but this law is not more conspirational than the law of conservation of angular momentum.

The correlation function needed in a simple model for Bell's Gedanken experiment, was calculated in Ref.\,\cite{GtHCA}. Our argument is similar to ones raised earlier, such as Ref.\,\cite{Vervoort-2013}.

\end{document}